\documentclass[11pt]{article}
\usepackage{amsmath,amsfonts}
\usepackage{tikz}
\usetikzlibrary{trees,er,snakes,shapes,mindmap}
\allowdisplaybreaks
\def \beq  {\begin{equation}}
\def \eeq  {\end{equation}}
\def \beqar {\begin{eqnarray}}
\def \eeqar {\end{eqnarray}}
\def\sqr#1#2{{\vcenter{\vbox{\hrule height.#2pt
\hbox{\vrule width.#2pt height#1pt \kern#1pt
\vrule width.#2pt}\hrule height.#2pt}}}}

\def\la {{\langle}}
\def\ra {{\rangle}}

\def\vf {{\varphi}}

\def\Tr {{\rm Tr}}

\def\bu {\bar{u}}

\def\bw {\bar{w}}

\def\vf {{\varphi}}

\def\del {\partial}

\def\D {{\cal D}}
\def\bz {{\bar{z}}}

\def\A {{\cal A}}

\def\H {{\cal H}}

\def\M{{\cal M}}

\def\half{\textstyle{1\over 2}}

\begin{document}
%%%%%%%%%%%%%%%%%%%%%%%%%%%%%%%%%%%%%%%%%%%%%%%%
%%%%%%%%%%%%%%%%%%%%%%%%%%%%%%%%%%%%%%%%%%%%%%%%
%%%%%%%%%%%%%%%%%%%%%%%%%%%%%%%%%%%%%%%%%%%%%%%%
%\fontfamily{pnb}\fontsize{12pt}{16pt}\selectfont
%\fontfamily{pzc}\fontsize{14pt}{16pt}\selectfont
%\fontfamily{pbk}\fontsize{12pt}{16pt}\selectfont
\fontfamily{cmr}\fontsize{11pt}{15pt}\selectfont
%\fontfamily{phv}\fontshape{ro}\fontsize{11pt}{14pt}\selectfont
%\fontfamily{ptm}\fontseries{m}\fontshape{r}\fontsize{12pt}{16pt}\selectfont
%\fontfamily{pnc}\fontseries{m}\fontshape{r}\fontsize{11pt}{15pt}\selectfont
%\fontfamily{ppl}\fontseries{m}\fontshape{r}\fontsize{11pt}{15pt}\selectfont
%\usefont{T1}{phv}{m}{it}
%%%%%%%%%%%%%%%%%%%%%%%%%%%%%%%%%%%%%%%%%%%%%%%
\def \CMP {{Commun. Math. Phys.}}
\def \PRL {{Phys. Rev. Lett.}}
\def \PL {{Phys. Lett.}}
\def \NPBProc {{Nucl. Phys. B (Proc. Suppl.)}}
\def \NP {{Nucl. Phys.}}
\def \RMP {{Rev. Mod. Phys.}}
\def \JGP {{J. Geom. Phys.}}
\def \CQG {{Class. Quant. Grav.}}
\def \MPL {{Mod. Phys. Lett.}}
\def \IJMP {{ Int. J. Mod. Phys.}}
\def \JHEP {{JHEP}}
\def \PR {{Phys. Rev.}}
\def \JMP {{J. Math. Phys.}}
\def \GRG{{Gen. Rel. Grav.}}
%%%%%%%%%%%%%%%%%%%%%%%%%%%%%%%%%%%%%%%%%%%%%%%
%%%%%%%%%%%%%%%%%%%%%%%%%%%%%%%%%%%%%%%%%%%%%%%
%%%%%%%%%%%%%%%%%%%%%%%%%%%%%%%%%%%%%%%%%%%%%%%
%%%%%%%%%%%%%%%%%%%%%%%%%%%%%%%%%%%%%%%%%%%%%%%
\begin{titlepage}
\null\vspace{-62pt} \pagestyle{empty}
\begin{center}
\rightline{CCNY-HEP-15/5}
\rightline{August 2015}
\vspace{1truein} {\Large\bfseries
Thermofield dynamics and Gravity
}\\
\vspace{6pt}
\vskip .1in
{\Large \bfseries  ~}\\
\vskip .1in
{\Large\bfseries ~}\\
\vspace{.6in}
{\large\sc V.P. Nair}\\
\vskip .2in
{\itshape Physics Department\\
City College of the CUNY\\
New York, NY 10031}\\
\vskip .1in
\begin{tabular}{r l}
E-mail:
&{\fontfamily{cmtt}\fontsize{11pt}{15pt}\selectfont vpn@sci.ccny.cuny.edu}
\end{tabular}

\fontfamily{cmr}\fontsize{11pt}{15pt}\selectfont
\vspace{.8in}
%\vspace{1.5in}%\vspace{0.3in}
%%%%%%%%%%%%%%%%%%%%%%%%%%%%%%%%%%%%%%%%%%%%%%%%
%%%%%%%%%%%%%%%%%%%%%%%%%%%%%%%%%%%%%%%%%%%%%%%%
\centerline{\large\bf Abstract}
\end{center}
Thermofield dynamics is presented in terms of a path-integral using coherent states, equivalently, using a coadjoint orbit action.
A field theoretic formulation in terms of fields on a manifold 
$\M \times {\tilde\M}$ where the two components have opposite orientation is also presented.
We propose formulating gravitational dynamics for noncommutative geometry using thermofield dynamics, doubling the Hilbert space modeling the noncommutative space.
We consider 2+1 dimensions in some detail and since
$\M$ and ${\tilde \M}$ have opposite orientation, the commutative limit
 leads to the Einstein-Hilbert action as the difference of two Chern-Simons actions.

\end{titlepage}

%%%%%%%%%%%%%%%%%%%%%%%%%%%%%%%%%%%%%%%%%%%%%%%%
%%%%%%%%%%%%%%%%%%%%%%%%%%%%%%%%%%%%%%%%%%%%%%%%
%%%%%%%%%%%%%%%%%%%%%%%%%%%%%%%%%%%%%%%%%%%%%%%%
%%%%%%%%%%%%%%%%%%%%%%%%%%%%%%%%%%%%%%%%%%%%%%%%
\pagestyle{plain} \setcounter{page}{2}
%%%%%%%%%%%%%%%%%%%%%%%%%%%%%%%%%%%%%%%%%%%%%%%%
%%%%%%%%%%%%%%%%%%%%%%%%%%%%%%%%%%%%%%%%%%%%%%%%
%%%%%%%%%%%%%%%%%%%%%%%%%%%%%%%%%%%%%%%%%%%%%%%%
%%%%%%%%%%%%%%%%%%%%%%%%%%%%%%%%%%%%%%%%%%%%%%%%
\section{Introduction}

It is well established by now that thermofield dynamics (TFD) gives a natural framework to analyze  time-dependent processes at finite temperature \cite{{TFD}, {vW-L}}.
 But the formalism goes beyond this limited context. It is ultimately  a method for describing mixed states as pure states in an enlarged Hilbert space.
Therefore, it is the natural theoretical framework for physical contexts where entropy plays an important role.
And nowhere is entropy more central or more mysterious than in gravity and so
it is not surprising that the description of states in a black hole background involves
thermofield dynamics \cite{{israel}, {maldacena}, {ads}}.

While most discussions of TFD focus on algebraic aspects of the formalism,
or on detailed Feynman diagram
techniques, in this paper we will write a general path-integral for TFD
in terms of functional integration over unitary matrices. For simplicity, we will consider finite-dimensional Hilbert spaces, with a large dimension taken at the end.
An alternate representation would be in terms of an auxiliary fermion field.
These representations of TFD are interesting in their own right,
but part of our  motivation is to apply this to gravity on noncommutative spaces
\cite{{general},{bal1}}.

The existence of entropy for empty gravitational backgrounds such as de Sitter space
suggests the idea of assigning a set of states to space itself.
In turn, this leads to the notion of noncommutative or fuzzy spaces,
since the basic premise of the latter is that some smooth manifolds can be obtained as 
an approximation to a Hilbert space of states as some parameter is taken to be very large.
(If we start with a finite-dimensional Hilbert space, this parameter is usually the dimension
itself.) The possibility that gravity might emerge from how this limit is taken was pointed out
 some time ago \cite{Nair2}. 
 What emerges naturally is Chern-Simons gravity. Here we will argue that
 one can get the Einstein-Hilbert action if the whole problem is set within TFD and
 we assign gravitational fields of opposite chirality to the physical system and the tilde system.
 In other words, our basic suggestion is that one must double the Hilbert space
modeling the noncommutative geometry and construct dynamics using thermofield dynamics.
 (The extension of TFD for fields and the related diagrammatic perturbation theory
 on a noncommutative space have been considered before \cite{amilcar}.
 This is different from what we propose here.)

There are, of course, many alternate approaches to gravity using
noncommutative spaces going back to the original suggestion by Connes, which leads to
the spectral action of Connes and Chamseddine \cite{connes}, 
Other approaches include 
the use of a $\theta$-deformed differential geometry, with and without Drinfeld twists \cite{{aschieri}, {bal}} and the emergence of spacetime and gravity from matrix models \cite{steinacker}.
(The matrix model approach has similarities to ours, but is still significantly different.)

There has also been more general interest in noncommutative spaces.
In fact, 
field theories on noncommutative spaces
have been an important topic of research for a long 
time now \cite{douglas}.
Such spaces can arise as brane solutions
in certain contexts
in string theory and in the matrix version of $M$-theory \cite{taylor}.
Gauge theories on such spaces can describe fluctuations of the brane solutions and this has also contributed to interest about field theories on fuzzy and noncommutative spaces.

Among noncommutative spaces, there is a subclass which can be described by finite-dimensional matrices; these are the fuzzy spaces and, by now, there are many examples of such spaces \cite{{general}, {bal1}}.
When the dimension of the matrices becomes large, these spaces tend to their smooth versions in terms of both the geometry and the algebra of functions on such spaces.
For much of what we do in this paper, we will use finite-dimensional Hilbert spaces, so
the discussion is within this class of spaces.
A more general starting point is possible, but we may note that
the finiteness of entropy for the de Sitter universe suggests that the use of a 
finite-dimensional Hilbert space to describe all degrees of freedom is worthy of consideration as a basic premise for physical theories.
We also point out another facet of this, namely, entropy considerations are also very suggestive of treating the event horizon as a fuzzy sphere \cite{dolan}.
The idea that entropy may play a stronger role has also been suggested recently, namely,
that gravity itself can be related to entropy and the first law of thermodynamics \cite{jacobson},
or from the first law for CFTs \cite{raamsdonk},
or that it may be an entropic force  \cite{verlinde}.
 There is also a recent approach based on a matrix model description of the ${\cal N} =4$ supersymmetric Yang-Mills theory which starts with a finite number of states, see \cite{herman}.

The time-evolution of a single quantum system starting from a given pure state
can be expressed by a path-integral using coherent states or in terms of unitary matrices.
The relevant action is in the form of a coadjoint orbit action \cite{bal3}.
In the next section, we rewrite TFD for a single quantum 
system in a similar way. This requires a doubled set of coherent states and the
action relevant for this path-integral
involves the trace over a matrix $P$ which has
the eigenvalue $+1$ for the system under consideration and $-1$ for the tilde system.
We can rephrase this in terms of an 
action defined on a closed contour on a cylinder
${\mathbb R} \times S^1$ with a single winding around the $S^1$ cycle. 
Multiple holonomies around
this cycle can be related to the R\'enyi entropy.

We then re-express this in a more familiar form as the functional integral of a field theory, the fields being two fermions. The fermions are
defined on two copies of a suitable K\"ahler manifold $\M$ with opposite orientation.
The fields are also coupled to 
a background field which is a multiple of the K\"ahler form 
on ${\cal M}$. A variant of this formulation is to take the fields to be
spinors and the action to be the massless Dirac action,.
A limit
 $c \rightarrow \infty$ has to be taken at the end, where $c$ plays the role of the speed of light in the action. One can also generalize this to a multipartite system
 by considering multiple copies of the
fields.

In section 3, we discuss how these ideas may apply to gravity, mostly in the setting of the
three-dimensional (or 2+1 dimensional) case. A doubling of the Hilbert space defining the fuzzy geometry is introduced. The basic (Euclidean) symmetry under consideration is $SU(2)_L \times SU(2)_R$. While the time-component of a gauge field for this symmetry appears in the thermofield action, the spatial components arise as auxiliary fields which give a simple way to encode the symmetry in the large $N$ limit. These  have to be eliminated, i.e., a specific choice
for these gauge fields has to be made, at the end.
One way to do this would be to choose them as extrema of the action in the large $N$ limit. Gravitational field equations arise as this choice.

One of the basic suggestions in this paper is that, for the doubled Hilbert space defining the geometry,
the gauge fields of, say, $SU(2)_L$ couple to one component while the gauge fields
of $SU(2)_R$ couple to the other, the tilde part. The action in the large $N$ limit becomes a Chern-Simons action for the $SU(2)_L$ fields and a similar one for the $SU(2)_R$ fields,
with a crucial negative sign, for the
tilde part.
The full action is thus the difference of two Chern-Simons actions, which is equivalent to
Einstein gravity in three dimensions.

We close with a brief discussion of the Minkowski signature and comments on comparison 
of the results with the literature. A short appendix elaborates on the Hall effect connection
for one version of the TFD path-integral.

\section{Generalizing thermofield dynamics}

\subsection{Action and functional integral for thermofield dynamics}

We start by considering the thermal average of an observable ${\cal O}$ at temperature
$\beta^{-1}$ defined by
\beq
\la {\cal O}\ra = \Tr (\rho \, {\cal O} ) = {1\over Z} \Tr \left( e^{-\beta H } {\cal O} \right),
\hskip .3in Z = \Tr \left( e^{-\beta H}\right)
\label{1}
\eeq
The density matrix $\rho$ corresponds to a mixed state. The basic idea of thermofield dynamics is to
represent the average $\la {\cal O}\ra$ as the expectation value of the operator
${\cal O}$ for a pure state.
This will require a doubling of the Hilbert space of states. It is easy to see that we cannot represent
$\rho$ as a pure state without doubling, since $\rho^2 = \rho$ for a pure state and
we have $\rho^2 \neq \rho$ for the thermal density matrix and no unitary transformation can change this property. If $\H$ denotes the Hilbert space of states (which can and will be taken to be finite dimensional for most of the discussion), then the Hilbert space for thermofield dynamics is
$\H \otimes {\tilde \H}$, where ${\tilde \H}$ is a copy of $\H$ itself.
A general state in $\H \otimes {\tilde \H}$ is of the form
$\vert m, {\tilde n}\ra$. The thermal vacuum is then defined as \cite{TFD}
\beq
\vert \Omega \ra = {1\over \sqrt{Z}}\, \sum_n e^{- \half \,\beta E_n } \, \vert n, {\tilde n}\ra
\label{2}
\eeq
We have used a basis of eigenstates of the Hamiltonian. In $\vert \Omega\ra$ 
the corresponding states from each Hilbert space contribute to this sum.
With this choice, it follows easily that
\beqar
\la \Omega \vert \, {\cal O}\, \vert \Omega \ra &=&
{1\over Z} \sum_{m, n} e^{- \half\,\beta ( E_n + E_m )}  \, \la m \vert {\cal O}\vert n\ra
\, \la {\tilde m}\vert {\tilde n}\ra\nonumber\\
&=& {1\over Z} \sum_n e^{-\beta E_n } \, \la n \vert {\cal O}\vert n\ra
= \Tr (\rho \, {\cal O})
\label{3}
\eeqar
where we have used the fact that ${\cal O}$ only acts on $\H$, and
$\la {\tilde m}\vert {\tilde n}\ra = \delta_{mn}$.
The thermal average is thus expressed as the expectation value over the pure state
$\vert \Omega\ra$. Thermodynamic entropy for $\rho$ will arise as an entanglement
entropy as we restrict the description to just one component, namely $\H$, of this
doubled Hilbert space.

The tilde Hilbert space is usually chosen as a copy of the dual space
of $\H$, namely $\H^*$. The motivation for this choice is that time-evolution
in ${\tilde \H}$ is then given by $-H$, with the full Hamiltonian being
\beq
\check{H} = H - {\tilde H} = H \otimes {\bf 1} - {\bf 1}\otimes H
\label{4}
\eeq
The state $\vert \Omega \ra$ obeys $\check{H} \vert \Omega \ra = 0$ and is independent of time.
Thus for a free bosonic field with single particle energies $\omega_k$,
\beq
\check{H} = \sum_k \omega_k \, ( a^\dagger_k a_k - {\tilde a}^\dagger_k {\tilde a}_k )
\label{5}
\eeq
It is possible to introduce a Bogolyubov transformation from $a, \, a^\dagger ,\, 
{\tilde a}, \, {\tilde a}^\dagger$  to $A, \, A^\dagger$ such that $\vert \Omega \ra$ is defined by
$A_k \vert \Omega \ra = 0$ \cite{vW-L}.
We will not go into this at this point.

A general density matrix $\rho$ obeys the Liouville equation
\beq
i \,  {\del \rho \over \del t} = H \, \rho - \rho\, H
\label{6}
\eeq
It is possible to construct an action whose variational equation of motion is
(\ref{6}). This is given by
\beq
S = \int dt\, \Tr \left[ \rho_0 \, \left(  U^\dagger \, i {\del U \over \del t} -
U^\dagger \, H\, U \right) \right]
\label{7}
\eeq
where $\rho_0$ is the initial density matrix and $U$ is a unitary matrix on $\H$.
The dynamical variable in (\ref{7}) is $U$ and $\rho$ is defined as
$\rho = U \, \rho_0 \, U^\dagger$. Notice that (\ref{7}) is in the form of a coadjoint orbit action.
It may be viewed as a general action for a general quantum system and is very useful in extracting
effective actions for collective phenomena \cite{sakita}. It is particularly suited for semiclassical and
large $N$ expansions. (A few papers which focus on this aspect, by no means an exhaustive list,
are \cite{{sakita}, {Nair1}}.)

Our first goal is to construct a similar action for the calculation of averages and correlators
in the thermofield state $\vert \Omega\ra$. Although we initiated the discussion
using  $\rho = Z^{-1}\, e^{-\beta H}$,
the idea of thermofield states can be used with any density matrix, so we will consider this more general situation. We start by introducing a different notation which will simplify the calculations.
As a basis for the Hilbert space, we introduce coherent states, representing the states
by wave functions $\phi_n (z)$, $\chi_n (w)$. 
We may view $\phi_n (z)$ (and likewise $\chi_n (w)$) as $U(N)$ coherent states for the case of
an $N$-dimensional Hilbert space; the coordinates $z$ and $w$ refer to the appropriate
coset space for such a construction.
Explicitly, there are many ways to construct the coherent states, for example, in terms of the
rank $n$ $SU(2)$ representations (with $N = n+1$), or rank $1$ representation of $U(N)$,
or using other subgroups of $U(N)$ if the dimensions are compatible. 
The coherent states may be viewed as sections of an appropriate line bundle over a suitable K\"ahler space $\cal M$ (of real dimension $2\,d$) (which would be ${\mathbb {CP}}^1 = SU(2)/ U(1)$ for 
the $SU(2)$ coherent states and ${\mathbb {CP}}^{N-1}$ for the rank $1$ U(N) coherent states).
In any case, for the coherent states, we can choose an orthonormal basis with
\beq
\int_{\cal M}  d\mu (\bz, z) \, \phi_n^* \, \phi_m = \delta_{nm}, 
\hskip .2in 
\int_{\cal M}  d\mu (\bw, w) \, \chi_n^*\, \chi_m = \delta_{nm}
\label{8}
\eeq
The thermofield state $\vert \Omega \ra$ can then be represented as
\beq
\vert \Omega \ra = \chi^*_n \, (\sqrt{\rho})_{nm} \, \phi_m
\label{9}
\eeq
Here we take $\sqrt{\rho}$ to act on the $\phi$'s, i.e., on $\H$. Thus $\sqrt{\rho}\, \phi$ is an element
of $\H$. The action of an operator ${\cal O}$ on $\vert \Omega\ra$ is given by
\beq
{\cal O} \, \vert \Omega \ra = \chi^\dagger \, \sqrt{\rho} \,\, {\cal O} \phi
\label{10}
\eeq
with the expectation value
\beqar
\la \Omega \vert \, {\cal O}\, \vert \Omega\ra &=&  \int (\phi^\dagger \sqrt{\rho}\, \chi)\,
(\chi^\dagger \sqrt{\rho}\,\, {\cal O} \phi )
= \int \phi^*_a  (\sqrt{\rho})_{ab} \, \chi_b\,
\chi^*_c\, (\sqrt{\rho})_{cd}  ({\cal O} \phi )_d\nonumber\\
&=& \Tr (\rho\, {\cal O})
\label{11}
\eeqar

Our notation also makes explicit some of the well known properties of TFD,
particularly its relation to $C^*$-algebras and the Tomita-Takesaki theory or the HHW version of the same \cite{HHW}.
 (For a concise review of this aspect of TFD, see
\cite{vW-L}.)  A key result is the existence of an antilinear operation $J$, referred to as
the modular conjugation, and a so-called modular operator
$\Delta$. These obey the properties
\beqar
\Delta^\dagger = \Delta , && \Delta > 0, \hskip .2in \Delta \, \vert \Omega \ra
= \vert \Omega \ra \nonumber\\
J^\dagger = J, && J^2 = 1, \hskip .2in J \, \vert \Omega \ra = \vert \Omega \ra
\label{12}
\eeqar
In our notation (\ref{10}) for $\vert \Omega\ra$, these are given by
$\Delta = e^{- \beta \check{H}}$ (for the thermal state $\rho = Z^{-1} e^{-\beta H}$),
and
\beq
J \, \phi = \chi^*, \hskip .3in J \, \chi = \phi^*, \hskip .3in
J \, \lambda \, \phi = \lambda^* \, \chi^*
\label{13}
\eeq
Thus
\beqar
J \, \vert \Omega \ra &=& J \, \left(\chi^\dagger \, \sqrt{\rho} \,\, {\cal O} \phi\right)
= \phi_a (\sqrt{\rho})^*_{ab} \, \chi^*_b = \chi^\dagger \sqrt{\rho^\dagger} \, \phi
= \chi^\dagger \sqrt{\rho} \, \phi\nonumber\\
&=& \vert \Omega \ra
\label{14}
\eeqar
where we used the fact that $\rho^\dagger = \rho$. All the properties
(\ref{12}) can be easily verified.

\begin{figure}[!b]
\begin{center}
\scalebox{1}{\includegraphics{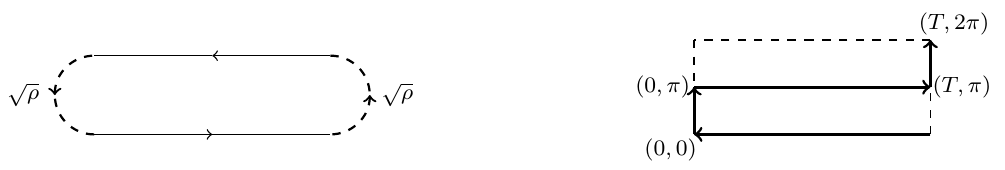}}
\caption{Time contours. The left side shows the contour for
equation (\ref{22}) and the right side for equation (\ref{new3}).}
\label{pic1}
\end{center}
\end{figure}

We now turn to the time-evolution of the states. For this purpose, we will rewrite $\Omega$
in a slightly different notation as
\beq
\Omega  (\bz, \bu ) = \sum_{n\, m} \psi_n(\bu ) \, (\sqrt{\rho})_{nm} \, \phi_m (\bz )
\label{15}
\eeq
Since we use the wave function representation for the states from now on, we write
$\Omega$ rather than $\vert \Omega \ra$. Also, we write
$\psi$ in place of $\chi^*$.
The time-evolved wave functions can be represented as a path integral,
\beq
\phi_n (\bz , t) = \int [\D z] \, e^{i\, S (z, \bz, t \,\vert z', \bz' )} ~
\phi_n (\bz', 0)
\label{16}
\eeq
where $ S (z, \bz, t \,\vert z', \bz' )$ is the action for the coherent states integrated from
$z'$, $\bz'$ to $z$, $\bz$ over time $t$. For the full thermofield state $\Omega$ we find
\beq
\Omega (\bz, \bu , t) = \int [\D z\, \D u] \, e^{i\, S (z, \bz, t \,\vert z', \bz' )} ~
e^{i\, {\tilde S} (u, \bu, t \,\vert u', \bu' )}~
\Omega (\bz' , \bu' ,0)
\label{17}
\eeq
Thus a vacuum-to-vacuum amplitude has the form
\beq
F = \int [\D z\, \D u] \,\, \Omega^* (z, u) ~e^{i\, S (z, \bz, t \,\vert z', \bz' )} ~
e^{i\, {\tilde S} (u, \bu, t \,\vert u', \bu' )}~
\Omega (\bz' , \bu')
\label{18}
\eeq
In the operator notation
\beq
\int \phi_k^*(z) ~e^{i\, S (z, \bz, t \,\vert z', \bz' )}~ \phi_l (\bz' ) = 
\la k \vert \, e^{-i H_z t } \, \vert l\ra
\label{19}
\eeq
where $H_z$ is the Hamiltonian for the $(z, \bz )$-system.
In other words, in the coherent state basis,
\beq
e^{i\, S (z, \bz, t \,\vert z', \bz' )} = \la z \vert \, e^{-i H_z t } \, \vert z' \ra
\label{19a}
\eeq
The amplitude $F$ can therefore be written as
\beq
F = \sum (\sqrt{\rho})^*_{kl} \, \la k \vert \, e^{-i H_z t} \, \vert a\ra
~\la l\vert \, e^{- i H_u t}\, \vert b\ra \, (\sqrt{\rho})_{ab}
\label{20}
\eeq
We choose $H_u = - H^T$, so that
\beq
F = \Tr \left( \sqrt{\rho}^\dagger \, e^{- i H t } \, \sqrt{\rho} 
\, e^{i H t} \right)
\label{21}
\eeq
If we introduce operators $A$, $B$ in the $(z, \bz)$-sector (or on $\H$), we can write,
for $t> t_1 > t_2$,
\beq
\la A(t_1) \, B(t_2) \ra = \Tr \left( \sqrt{\rho}\,\, U(t, t_1) \,A \, U(t_1 , t_2) \, B\, U(t_2, 0)
\, \sqrt{\rho} \,\, U^\dagger (t, 0)\right)
\label{22}
\eeq
In terms of a time-contour, this may be represented as shown
in Fig.\,\ref{pic1}. This is not the more familiar
Schwinger-Bakshi-Mahanthappa-Keldysh (SBMK) closed time contour \cite{SBMK}.
That has insertion of $\rho$ at $t =0$ and identity at $t =T$.
%%%%%%%%%%%%%%%%%%%%%%%%%%%%%%%%%%%%
%%%%%%%%%%%%%%%%%%%%%%%%%%%%%%%%%%%%
%%%%%%%%%%%%%%%%%%%%%%%%%%%%%%%%%%%%
%%%%%%%%%%%%%%%%%%%%%%%%%%%%%%%%%%%%
For equilibrium cases, it is equivalent to the $F$ given here.
If $\rho$ is not the equilibrium choice, then there are several amplitudes (with different 
physical meanings) one can consider.
We can define a more general
state specified by a matrix $K$ as
\beq
\Omega_K = \sum_{n\, m} \psi_n (\bu ) K_{nm} \, \phi_m (\bz )
\label{23}
\eeq
The SBMK contour is then obtained for the amplitude
\beq
F_{1\,\rho} = \int [\D z\, \D u] \, \, \Omega_{\bf 1}^* (z, u) ~e^{i\, S (z, \bz, t \,\vert z', \bz' )} ~
e^{i\, {\tilde S} (u, \bu, t \,\vert u', \bu' )}~
\Omega_\rho (\bz' , \bu')
\label{24}
\eeq
This is the physically relevant amplitude (with insertions of operators such as
$A$, $B$ as required) for time-evolution of a closed statistical system
starting with a given $\rho$. However, if we view the physical system under consideration as 
a subsystem within a larger closed system which is in a pure state
which is an eigenstate of the total Hamiltonian, 
then the amplitude (\ref{22}) would be the relevant one.

We now turn to the action which will give the expected behavior for the $(z, \bz)$- and
$(u, \bu)$-sectors. It is given by
\beq
S = \int dt \left[ \left( i \, \bz_k {\dot z}_k - \bz_k \, H_{kl} z_l \right)
+ \left( i \, \bu_k {\dot u}_k + \bu_k \, H^T_{kl} u_l \right) \right]
\label{25}
\eeq
with the constraints
\beq
\bz_k \, z_k = 1, \hskip .3in \bu_k \, u_k = 1
\label{26}
\eeq
This action is easily quantized in terms of geometric quantization.
For the $(z, \bz)$-sector, the canonical one-form and two-form are given by
\beq
A = {i \over 2} ( \bz_k \, dz_k - d\bz_k\, z_k ) , \hskip .3in
\omega = i \, d\bz_k \wedge dz_k
\label{27}
\eeq
The polarization condition on the wave functions can be chosen as
\beq
\nabla_z \, \Psi = \left( {\del \over \del z_k } + {\bz_k \over 2} \right) \, \Psi = 0
\label{28}
\eeq
which leads to wave functions of the form
\beq
\Psi = e^{- z_k \bz_k /2} \, f (\bz )
\label{29}
\eeq
with $z_k$ acting as $\del /\del \bz_k$ on the the $f$'s.
The constraint shows that the $f$ can have one power of $\bz$, 
which implies that $f (\bz) \sim \bz_k$.
There are exactly $N$ states, giving the rank $1$ representation of $U(N)$.
The Hamiltonian operator is
\beq
H = \bz_k \, H_{kl} {\del \over \del \bz_l} 
\label{30}
\eeq
We see that matrix elements of this Hamiltonian reproduce $H_{kl}$.
For the $(u, \bu)$-sector, we again have $\Psi = \exp (- u\cdot\bu /2) \, f(\bu)$ with
$u_k$ acting as $\del/ \del\bu_k$. The Hamiltonian is
\beq
H = - \bu_k \, H^T_{kl} \, {\del \over \del \bu_l}, 
\hskip .3in \la k\vert \, H \, \vert l\ra = - H^T_{kl} 
\label{31}
\eeq
The operation $H \rightarrow - H^T$ represents charge conjugation in the Lie algebra of $U(N)$.
It is useful to define
\beq
z_k = \xi_{k1}, \hskip .2in \bu_k = w_k = \xi_{k 2}, \hskip .3in 
P = \left[ \begin{matrix} 
1&0\\ 0&-1\\ \end{matrix} \right]
\label{32}
\eeq
With a partial integration on the term involving the time-derivative of $u_k$ in (\ref{25}), 
we can bring the action to the form
\beqar
S &=&\int dt~ \sum_{\alpha, \beta = 1,2} P_{\alpha\beta}\left(  i {\bar\xi}_{k \beta} {\dot \xi}_{k \alpha} 
- {\bar \xi}_{k\beta} \, H_{kl} \, \xi_{l \alpha} \right)\nonumber\\
&=& \int dt~ \Tr \left[ P \, \left( i\, \xi^\dagger \, {\dot \xi} - \xi^\dagger \, H \, \xi \right)\right]
\label{33}
\eeqar
The variables $z_k, \, u_k$ with the constraint (\ref{26}) define (two copies of)
$\mathbb{CP}^{N-1}$, so it is useful to 
use a notation in terms of group elements.
Writing $\xi_{k\alpha } = U^{(\alpha)}_{k 0} = \la k \vert U^{(\alpha)} \vert 0\ra$, for 
two unitary matrices $U^{(\alpha)}$, we find
\beqar
S &=& \int dt~ \left[  \left(  i\,U^{(1)\dagger}  {\dot U^{(1)}} - U^{(1)\dagger} H\, U^{(1)} \right)_{00}
-  \left(  i\,U^{(2)\dagger}  {\dot U^{(2)}} - U^{(2)\dagger} H\, U^{(2)} \right)_{00}\right]
\label{34}
\eeqar
The $N$ states forming the fundamental representation of $SU(N)$ can be viewed as
being generated from a highest weight state, usually the vacuum, by the action of various operators.
Group theoretically, they can be constructed from the elements $\la  k\vert U \vert  0\ra$
for a chosen state $\vert 0\ra$. We have explicitly included this in our notation.
The action (\ref{34})  is close to the form (\ref{7}), but the key difference is that
it is the difference of two actions of the form (\ref{7}).
In the same notation, the state
$\Omega$ can be written as
\beq
\Omega = \bz_k \, \sqrt{\rho}_{kl} \, w_l = {\bar \xi}_{k1}\, \sqrt{\rho}_{kl}\,\, \xi_{l2} =
 \la 0\vert U^{(1)\dagger} \,\sqrt{\rho}\, \,U^{(2)} \vert 0\ra
\label{35}
\eeq
Although we have used coherent states to arrive at (\ref{34}) and (\ref{35}),
these results are no longer dependent on any specific representation for the states.
In terms of (\ref{35}), we can represent (\ref{18}) as
\beq
F = \int [{\cal D} U] ~ e^{i S }~ \la0\vert U^{(2)\dagger }(t)) \,\sqrt{\rho^\dagger}\, \,U^{(1)} (t) \vert 0\ra\,
\la 0\vert U^{(1)\dagger} (0) \,\sqrt{\rho}\, \,U^{(2)}(0) \vert0\ra
\label{35a}
\eeq
with $S$ as given in (\ref{34}) and $[\D U]$ denotes functional integration 
for the two copies of $\mathbb{CP}^{N-1}$.

In calculating $F$ as in (\ref{18}), or correlators as in (\ref{22}), effectively we have a closed time-contour with insertions of $\Omega$ at two points corresponding to $t = 0$ and $t = T$
with $T\rightarrow \infty$ eventually.
This suggests a neat way to rewrite these results. We will use a complex time variable
$\tau = t + i \, \theta$, with the identitfication $\theta \sim \theta + 2 \pi$, so that $\tau$ describes
a cylinder ${\mathbb R} \times S^1$. Further we introduce a one-form
\beq
{\mathcal A} = - i \left[ \, H \, dt + {i\over 2 \pi} \log \rho \, d\theta \right]
\label{new1}
\eeq
The contour of integration $C$ is then taken as starting at $(t, \theta )  = (0,0)$ going to
$(0, \pi)$ to $(T, \pi)$ to $(T, 2\pi ) \sim (T, 0)$ to $(0,0)$, as shown in the second
figure in Fig.\,{\ref{pic1}}. This goes around $S^1$ once. We then consider
the ${\mathcal P} e^{\oint_C \A}$, where $\mathcal P$ denotes path-ordering along $C$.
For the segments along the $t$-direction, we can use (\ref{19a}) to obtain the action as before.
For the
segments $(0,0)$ to $(0, \pi)$ and $(T, \pi) $ to $(T, 2 \pi )$ 
along the imaginary time-direction, we get factors of
$\sqrt{\rho}$. 
For this, we use the same result as (\ref{19a}) with $(i/2) \log \rho$ in place of the Hamiltonian
$H$. Thus
\beq
e^{i S (w, z', \pi )} = \la w \vert \, e^{\half \log \rho} \,\vert z'\ra
\label{new2}
\eeq
Notice that the factor $\int U^{\dagger} {\dot U} \, d\tau$ is insensitive to
whether we integrate along real or imaginary directions. We also take $H$ to be independent of
$\theta$ so that $H ({\rm at}~\theta =  \pi ) = H ({\rm at}~ \theta = 0)$. With this result, we can write
the path integral for thermofield dynamics as
\beq
F_J = \int [{\mathcal D} U] ~ \exp \left[ \oint_C \left( - U^\dagger \, {\dot U} 
+ U^\dagger \, {\mathcal A} \, U \right)_{00}  + \oint AJ + B J' \right] 
\label{new3}
\eeq
where we have also introduced sources which facilitate the insertion of operators
$A$, $B$, etc. Setting $J = J' =0$ gives the vacuum to vacuum transition amplitude.
In operator notation, this is given by
\beq
F (C) = \Tr \, {\mathcal P} e^{ \oint_C {\mathcal A}}
\label{new4}
\eeq
Thus $F(C)$ is the Wilson loop integral for the connection
(\ref{new1}) taken along the contour $C$ which winds once around the $S^1$ direction
of the cylinder.
Consider now a small deformation of the contour $C \rightarrow C'$ at a point
$\tau_0$ along the path.
Evidently
\beq
F(C') - F(C) = \Tr \, {\mathcal P} \exp\left({\int_{\tau_0}^{(0,2\pi)} }\right) \, {\cal F} \sigma~
{\mathcal P} \exp\left( {\int^{\tau_0}_{(0, 0)} }\right)
\label{new5}
\eeq
where $\sigma = \delta t \wedge \delta \theta$ is the infinitesimal area
between $C'$ and $C$ at $\tau_0$. The curvature ${\cal F}$ is given by
\beq
{\cal F} = {1\over 2 \pi} {\del \over \del t} \log \rho  + i \, {\del H \over \del \theta}
+ {i \over 2 \pi} \, [ \log \rho , H ] 
\label{new6}
\eeq
In our case, $\del H / \del \theta \, = 0$ and ${\del \rho /\del t} = 0$. 
(We have insertions of the same $\rho$ at $t =0$ and $t= T$.)
Thus, the curvature $\cal F$ vanishes if $[ \rho, H ] = 0$.
For the equilibrium case where this is obtained, we can do deformation of the contour
$C$ reducing $F(C)$ to a simple holonomy  
of the connection
(\ref{new1}) around the $S^1$-direction of the cylinder ${\mathbb R}\times S^1$.
(This is without insertions of operators; for correlators there will be a nontrivial segment along
the $t$-direction.)

We can go further and consider multiple windings around the $S^1$ component of the
cylinder. We can consider windings inserted at any value of $t$, defining
$ W(C, n, t) = F(C+ C_n (t))$ where $C+ C_n (t)$ corresponds to the contour $C$ with
extra $n$ windings around $S^1$ at the point $t $.
Thus the R\'enyi entropy is defined in terms of the holonomy
as 
\beq
S_R (t) = - \left( {W(C, n, t)  - 1\over n -1} \right)
\label{new7}
\eeq
Since thermofield dynamics uses pure states, entropy must be defined in terms of the basic
observables of the formalism. For the vacuum amplitude, i.e., in the absence of insertions of operators on the real time line,
the holonomy is the only observable we have and hence
 equation (\ref{new7}) gives the natural definition.
 
 \subsection{A field theoretic representation}
 
 We have represented thermofield dynamics in (\ref{35a}) as a functional integral 
 over group elements. There are other equivalent ways of representing it, one of which is as
 the functional integral of a field theory. This is will prove useful as well. The basic idea is that the quantum system can be viewed as the one-particle sector of a field theory. The time-evolution matrix element (\ref{19}) can be written as
 \beqar
 \la k \vert e^{-iHt} \vert l\ra &=& \la 0\vert a_k \, e^{-i H t} \, a^\dagger_l \vert 0\ra
 = \la 0\vert T\, a_k(t) \, a^\dagger_l (0) \vert 0\ra\nonumber\\
 &=& {\cal N} \, \int [da \,da^*] ~ e^{i S} \,\, a_k(t) \, a_l^\dagger (0)
 \label{new8}\\
 S&=& \int dt \left[ a_k^* (i \del_0 ) a_k - a_k^* \, H_{kl} \, a_l \right]
 \label{new9}
 \eeqar
 and $\cal N$ is the standard normalization factor,
 \beq
{\cal N}^{-1} = \int [da \ da^*] ~ e^{i S}
\label{new10}
\eeq 
Further, we can introduce a $(z, \bz)$-dependent field (on ${\cal M}$)
given by $\psi (z, \bz, t) = \sum_k a_k \, z_k$,
$\psi^\dagger (z, \bz, t) = \sum_k a^\dagger_k \, \bz_k$.
(We use $\psi^*$ when we integrate over the c-number versions of these
fields in the functional integral.)
The diagonal coherent state representation of operators also allows us to introduce
$H(z, \bz)$ such that
\beq
H_{kl} = \int_{\cal M} d\mu (z, \bz)~ \bz_k \, H(z, \bz) \, z_l
\label{new11}
\eeq
The action (\ref{new9}) can thus be written as
\beq
S = \int dt\,d\mu(z,\bz)~ \left[ \psi^* (i \del_0 ) \psi - \psi^\dagger \, H(z, \bz) \, \psi
\right]
\label{new12}
\eeq
The part of the action (\ref{33}) with the negative eigenvalue for $P$ can be represented 
in a similar way, with a field $\phi, \phi^\dagger$.

The fields $\psi$, $\psi^*$ (and $\phi, \, \phi^*$) are restricted since
$\psi$ is holomorphic and $\psi^*$ is antiholomorphic.
Thus the functional integration over these fields have also to be suitably restricted.
A convenient way to extend the functional integration over all fields and still restrict the dynamics to
holomorphic $\psi$'s is to use the Landau level trick.
We take $\psi$ to describe a charged particle on $\cal M$ with a background 
magnetic field which is constant in a suitable basis. This can be done for
by taking the field to be proportional to the K\"ahler two-form $\omega$ on $\cal M$ and consider the action
\beq
S = \int dt\,d\mu(z,\bz)~ \left[ \psi^* \left(i\, \del_0  -  \, A_0(z, \bz)  + {D^2 +E_0 \over 2 m}\right)\, \psi
\right]
\label{new13}
\eeq
where $D_i = \nabla_i -i A_i$ is the gauge and Levi-Civita covariant derivative on $\psi$ and
$d (A_i dx^i) = n\, \omega$ for some parameter $n$.
 $E_0$ is the lowest eigenvalue of
$-D^2$. Further, $A_0 (z, \bz) = H(z, \bz)$ is the Hamiltonian of the original theory. 
$m$ is a parameter which we will take to be zero eventually.
The Hamiltonian for $\psi$ is $H' + A_0$, with
\beq
H' = - {D^2 +E_0\over 2 m} 
\label{new14}
\eeq
The eigenstates of $H'$ are the Landau levels,the lowest of which obeys a holomorphicity condition, and has zero eigenvalue since we subtracted $E_0$ (which is the lowest eigenvalue of
$-D^2$). The higher states are not holomorphic,
but will decouple as $m \rightarrow 0$. The dynamics will thus be restricted to
the lowest state which corresponds to holomorphic $\psi$'s.
Finally, we introduce
\beq
\Omega (\psi^* , \, \phi^*) =
\int_{\cal M} d\mu (z,\bz) d\mu(w, \bw)~ \psi^* (z)\, \phi^* (w) 
\, z_k \sqrt{\rho}_{kl} \, w_l
\label{new15}
\eeq
Collecting these results together, we conclude that thermofield dynamics is given by
\beqar
F &=&{\cal N}  \int [d \psi d\psi^*\, d\phi d\phi^*]~ e^{iS} ~\Omega^* (t) \, \Omega (0)
\label{new14}\\
S&=&\int dt\,d\mu(z,\bz)~ \biggl[ \psi^* \left(i\, \del_0  -  \, A_0(z, \bz)  + {D^2 +E_0 \over 2 m}\right)\, \psi 
\nonumber\\
&& \hskip 1in - ~ \phi^* \left(i\, \del_0  -  \, A_0(z, \bz)  + {D^2 +E_0 \over 2 m}\right)\, \phi  
\biggr]\nonumber\\
&=&\int dt \int_{\cal M} d\mu(z,\bz)~ \psi^* \left(i\, \del_0  -  \, A_0(z, \bz)  + {D^2 +E_0 \over 2 m}\right)\, \psi 
\nonumber\\
&& \hskip .2in + \int dt \int_{\tilde{\cal M}} d\mu (z, \bz)~ \phi^* \left(i\, \del_0  -  \, A_0(z, \bz)  + {D^2 +E_0 \over 2 m}\right)\, \phi  
\label{new16}
\eeqar
As usual, 
${\cal N}^{-1}$ is given by the integral of $e^{i S}$ over all fields.
In the second line of (\ref{new16}) we take ${\tilde {\cal M}}$ to have the orientation opposite to that of
${\cal M}$.
The fields may be taken to be bosonic, but it will turn out to be more convenient to take them as
fermionic fields.

A simple extension will give a generalization of this result to a multipartite system
with identical components. Going back to
(\ref{new9}), we take the states to be 
of the form $\vert k\ra = \vert \alpha \, I\ra \in \H_1 \otimes \H_2$ and define
a set of fermion fields $\psi_I = \sum_\alpha a_{\alpha I} \, z_\alpha$.
The action may now be written as
\beqar
S&=&\int dt \int_{\cal M} d\mu(z,\bz)~ \psi^*_I  \left(i\, \del_0  \delta_{IJ} -  \, (A_0(z, \bz)_{IJ}  + {D^2 +E_0 \over 2 m} \delta_{IJ} \right)\, \psi_J
\nonumber\\
&& \hskip .2in + \int dt \int_{\tilde{\cal M}} d\mu (z, \bz)~ \phi^*_I \left(i\, \del_0 \delta_{IJ} -  \, (A_0(z, \bz))_{IJ}  + {D^2 +E_0 \over 2 m}\delta_{IJ}\right)\, \phi_J
\label{new17}
\eeqar
We may interpret the labels $I, J$ as corresponding to some internal symmetry or 
degrees of freedom.

There is one more improvement we can do on this formula.
If ${\cal M} \times {\mathbb R}$ admits spinors, we can replace
the action by the Dirac type action
\beq
S = \int dt \int_{\cal M} d\mu (z, \bz)~ {\bar \Psi}_I ( i \gamma^\mu D_\mu )_{IJ}  \Psi_J
+ \int dt \int_{\tilde{\cal M}} d\mu (z, \bz)~ {\bar \Phi}_I ( i \gamma^\mu D_\mu )_{IJ} \Phi_J
\label{new18}
\eeq
where $\Psi$ and $\Phi$ are spinors, $\gamma^\mu$ are the standard Dirac matrices and
${\bar\Psi} = \Psi^\dagger \gamma^0$, ${\bar\Phi} = \Phi^\dagger \gamma^0$.
The Hamiltonian for $\Psi/ \Phi$ now has the form
$H' +A_0$ with $H' = -i \gamma^0 \gamma^i D_i$.
The eigenstates of $H'$ are again Landau levels; there are zero modes
for $H'$ which satisfy a holomorphicity condition.
The other levels are separated by a gap of order of the magnetic field $\sim c\,n\, \omega$, where
$c$ plays the role of the speed of light for the action (\ref{new18}), relating $\del/\del t$ and
$\del/ \del x$. Taking $c \rightarrow \infty$ all the nonzero eigenstates decouple\footnote{This is not the usual nonrelativistic limit
since we have set the mass to zero in (\ref{new18}).}.
(We assume, as usual, that the negative energy levels are all filled.)
The degeneracy of states is controlled by $n \, \omega$ and remains finite in this limit.
There is also another limit we can take, namely, $n \rightarrow \infty$; in this case, not only the nonzero levels decouple, the degeneracy of the zero modes also tends to infinity.
This is equivalent to an expansion in inverse powers of the background field.

\subsection{A quick summary}

A quick recapitulation of the results in this section will be useful. We have expressed thermofield dynamics for a single quantum system as a functional integral over a unitary group, the action for which is given in (\ref{34}). This action involves the trace over a matrix $P$ which has
the eigenvalue $+1$ for the system under consideration and $-1$ for the tilde system.
Equivalently, one can use an action defined on a closed contour on a cylinder
${\mathbb R} \times S^1$ with one winding around the $S^1$ cycle. Multiple holonomies around
this cycle can be related to the entropy.

One can also express thermofield dynamics  as field theoretic functional integral
with two fields defined on two copies of a suitable K\"ahler manifold with opposite orientation.
The fields are subject to a background field which is a multiple of the K\"ahler form 
on the manifold ${\cal M}$. The fields can be taken to be spinors and the action to be the massless Dirac action,
with a limit $c \rightarrow \infty$ at the end, where $c$ plays the role of the speed of light in the action.
Additional degrees of freedom can be incorporated  by considering multiple copies of the
fields, with, generally, a nonabelian symmetry acting on them.

\section{Gravity from noncommutative spaces}

A particularly interesting situation to which the foregoing analysis can be applied
is a formulation of gravity using noncommutative or fuzzy spaces.
The relevance of fuzzy spaces starts with the question: What if we quantized gravity?
In that case, we would have a  Hilbert space of states, and the continuous manifold
description would be obtained as an approximation for large number of degrees of freedom.
This is clearly the realm of noncommutative geometry or fuzzy spaces.

A scenario for implementing this idea would be as follows.
We can model noncommutative spaces
in terms of the lowest Landau level of a quantum Hall system
\cite{{Szabo}, {NC-LLL}}.
For example, we can think of the complex projective space
$\mathbb{CP}^k$ as $SU(k+1)/ U(k)$. It is thus possible to consider uniform background fields
on $\mathbb{CP}^k$ which are valued in the algebra of $U(k)$ and which are proportional to the curvatures of $\mathbb{CP}^k$. Single particle wave functions in this background will fall into
$SU(k+1)$ multiplets. The lowest such set of states can be represented by holomorphic wave functions. Put another way, these wave functions are sections of a holomorphic $U(k)$ bundle
on $\mathbb{CP}^k$. They form an $N$-dimensional Hilbert space
$\H_1$ which may be viewed as a model for the fuzzy version $\mathbb{CP}_F^k$ of 
the differential manifold ${\cal M} = \mathbb{CP}^k$. Functions on the fuzzy space are $N \times N$ matrices viewed as linear transformations on $\H_1$.
In the large $N$ limit, we recover the usual commutative algebra under pointwise multiplication
of functions on $\mathbb{CP}^k$. This limit can be analyzed by using classical functions
(on $\mathbb{CP}^k$) to represent operators and using $*$-products to
represent operator products.

The isometry group for $\mathbb{CP}^k$ is $SU(k+1)$. Hence, background fields valued in
the algebra of $U(k)$ amount to connections and curvatures for (at least part of) the isometry group.
Thus these background fields, since they correspond to gauging the isometries,
 can be viewed as describing gravitational degrees of freedom.
An action for these fields, which may be derived from (\ref{new17}) or
(\ref{new18}) (or from (\ref{36}) given in the appendix), would thus be a gravitational action.

This point of view regarding gravity was suggested many years 
ago \cite{Nair2}. The action
used in that case was (\ref{7}) or (\ref{38}), not the thermofield case with both positive and negative eigenvalues for $P$. Simplifying it using $*$-products of $\mathbb{CP}^k$ led to the Chern-Simons term for the $\underline{U(k)}$-valued gauge fields as the leading term in the action. 
However, there were several points which were not clear at that stage.
{\it A priori},
since we start with $\H_1$, the choice of modeling this as $\mathbb{CP}_F^k$ is arbitrary.
The leading CS term which emerges in the large $N$ limit, being topological, 
is not sensitive to the metrical details of $\mathbb{CP}^k$, but the subleading ones are.
Further, for gravity on a $2k$-dimensional spatial manifold ($+$ time), we need gauge fields valued in the algebra of the corresponding Poincar\'e group, 
(or de Sitter group, including a cosmological constant), with the gauge fields corresponding to the
translations being the frame fields and those for the Lorentz transformations giving the spin connection.
Thus, with Euclidean signature, we need $SO(2k +2)$ rather than $U(k)$. Finally, it was not clear
how one could get Einstein gravity rather than CS gravity. With the thermofield approach,
one can improve on some of these problems.

In this paper, we make two basic suggestions. The first is that in discussing
gravity using the large $N$ limit of a Hilbert space $\H_1$, we should use thermofield dynamics.
This is motivated by the well-founded expectation that entropy should play 
an important role in gravity and that thermofield dynamics, which can incorporate entropy within a formalism of pure states, is therefore a natural framework. We thus double the Hilbert
space to $\H_1 \otimes \H_1^*$. 
The large $N$ approximation of these spaces by a manifold $\M$ will introduce gauge fields
corresponding to the frame fields and spin connection of $\M$.
Our second suggestion is that these gauge fields 
for the physical system and the tilde system should be considered as parity conjugates of each other. This is related to the fact that the orientation of the manifold used to define the states, say in (\ref{new17}) or (\ref{new18}),
is reversed for the tilde system.
Equivalently, this emerges from the matrix $P$
in (\ref{36}). With this choice, the action, as we will see below, leads to Einstein gravity rather than Chern-Simons gravity.

Some of the key ingredients are then the following.
\begin{enumerate}
\item We need a Hilbert space $\H$ which carries a representation of a group $G$. The latter will eventually become the isometry group of the continuous spacetime ${\cal M} \times \mathbb{R}$
which will emerge as we take the limit of some parameter $\theta \rightarrow 0$. ($\theta$ could 
be $\sim N^{-\alpha}$, for some
$\alpha > 0$.) $G= U(k)$ for the discussion given above, but could be more general,
and, in fact, will be $SO(4)$ for the case of three-dimensional Euclidean gravity.

\item The Hilbert space will have three components, $\H_1 \otimes \H_2 \otimes \H_3$
with states of the form $\vert \alpha, a, I\ra$, where $\H_3$ refers to any matter system
of interest.
For gravity, as a first approximation, we will not need to consider excitations of the matter system, which means that we can restrict the matter fields to the ground state.  In this case, the
states in $\H$ will be taken to be of the form $\vert \alpha , a, 0\ra$ corresponding to a representation
$R_1 \otimes R_2$ of $G$ with the transformation
\beq
\vert \alpha, a,  0\ra' = g^{(1)}_{\alpha \beta} \, g^{(2)}_{ab} \, \vert \beta, b, 0\ra
\label{40b}
\eeq
We will consider $R_2$ to be a fixed representation, and take the dimension of $R_1$ to be very large to approximate $\H_1$ by a continuous manifold. In the case of an infinite dimensional representation $R_1$, we take a limit of the highest weight vector to approximate to a continuous manifold. $R_1$ must be a highest weight representation to define symbols and $*$-products.
These representations have to be unitary as required by quantum mechanics.

It is possible to reduce the product $R_1 \otimes R_2$ in terms of irreducible representations but this will not be important for us.
({\it A priori}, one could consider different groups $G_1$ and $G_2$ acting
on $\H_1$ and $\H_2$ respectively. We have no good argument to exclude this,
except that the minimal way is to use the same group.)
\item To define the functions representing operators
 and $*$-products, we will need a set of wave functions. These will be 
obtained by quantizing the K\"ahler space $G/H$ for some suitable $H \in G$. This will also define
$R_1$ (and $\H_1$). Gauge fields will emerge as part of the the
procedure for the large $N$ limit.
\end{enumerate}
Given this structure, it is possible to simplify the trace of an operator, say, $\Tr (i H)$. It gives
a Chern-Simons form of the appropriate dimension. For a spatial manifold 
$M$ of dimension $2k$, the action (\ref{36}) will simplify as the integral of a Chern-Simons 
$(2k+1)$-form with gauge group $G$, more precisely as the difference of
Chern-Simons actions for the two chiralities.
The gauge fields on the continuous spacetime are introduced to express the unitary transformations on $\H$ in terms of
$G$-transformations on the continuum ${\cal M}\times \mathbb{R}$.
In a sense, these gauge fields define the small $\theta$ (or large $N$) limit we are taking. The natural question then is whether we can choose them in some optimal fashion. As the optimization requirement, we extremize the limit of the action (\ref{36}) with respect to $A_\mu$. These 
optimization conditions are to be considered as the equations of motion for gravity.

Equivalently, we can use the field theoretic action (\ref{new18}) with the spinor fermionic
fields of the form
$\Psi_{i , 0}(z, \bz, t)$, $\Phi_{i , 0}(z, \bz, t)$. 
The gauge fields are matrices of the form
\beq
A_{ab} = A_i dx^i\, \delta_{ab}  + (A_\mu)_{ab} dx^\mu
\label{40c}
\eeq
where $d(A_i dx^i) = n \omega$ and $(A_\mu)_{ab}$
are valued in the Lie algebra
of $G$; they are matrices in the representation
$R_2$.

We now show how these ideas can be applied to gravity in three dimensions.
In three Euclidean dimensions, it is well known that
Einstein gravity can be formulated as a Chern-Simons theory \cite{3dgrav}.
This is true for Minkowski signature as well, but we will first discuss the Euclidean case.
For this, we will use
the generators of $SO(4)$ in the spinor representation, with $4\times 4$ hermitian
$\gamma$-matrices.
The translation generators $P_a$ and the rotation generators $S_{ab}$ are defined as
\beq
P_a = {\gamma_3 \, \gamma_a \over 2i l}, \hskip .3in
S_{ab} = {1\over 4i} \left( \gamma_a \gamma_b - \gamma_b \gamma_a \right), \hskip .2in 
a, b = 0, 1, 2.
\label{42}
\eeq
Here $l$ is a quantity with the dimensions of length which is related to the cosmological constant.
The action is then given by
\beq
S = - {l \over 32 \pi G} \int \Tr \left[ \gamma_5 \, \left( A \, dA + {2\over 3}
A^3 \right)\right]
\label{43}
\eeq
The gauge field $A$ is built of the frame field $e^a= e^a_\mu\, dx^\mu$ and the
spin connection $\omega^{ab} = \omega^{ab}_\mu \,dx^\mu$ as
\beq
A = -i \, P_a \, e^a - {i \over 2} \omega^{ab} \, S_{ab}
\label{44}
\eeq
The trace has the property that it gives a pairing between the $P_a$ and $S_{bc}$ only,
\beq
\Tr ( \gamma_5 \, P_a \, S_{bc} ) = {1 \over l} \,\epsilon_{abc}
\label{45}
\eeq
This pairing, rather than the usual Cartan metric of $SO(4)$, is crucial in being able to write
Einstein gravity as a Chern-Simons theory. In fact, simplifying (\ref{43}) using
(\ref{44}, \ref{45}) gives
\beq
S= {1\over 16 \pi G} \int d^3x \, \det e~ \left( R - { 3 \over 2 l^2} \right)
\label{46}
\eeq
Since the $SO(4)$ generators split into two chiral $SO(3)$'s corresponding to the subspaces
with $\gamma_5 = 1$ and $\gamma_5 = -1$, i.e., $SO(4) \sim SO(3)_L \times
SO(3)_R$, we see that the pairing (\ref{45}) can indeed be reproduced by the thermofield action is we take $P$ to be proportional to $\gamma_5$.
This is precisely what we propose to do.

We will now show how this action can be obtained from thermofield dynamics.
For this purpose, we can use the formulation
(\ref{new18}) in terms of spinor fields, but it is more illuminating to first
see how the action (\ref{43}) or (\ref{46}) emerges from (\ref{36}) in the large $N$ limit.
The $N$-dimensional space $\H_1$ can be taken to correspond to
a representation of $SU(2)$.
Since $SU(2)/U(1) = \mathbb{CP}^1$, we can think of these states as describing
fuzzy $\mathbb{CP}^1$. The states are thus of the form
$\vert \alpha , i\ra$, where for $i = 1,2 $, we have an action of $SU(2)$
given by
\beq
\vert \alpha , i\ra \rightarrow h^{(s)}_{\alpha\beta} \, h_{ij} \, \vert \beta, j\ra
\label{47}
\eeq
where $h^{(s)}_{\alpha\beta}$ denotes the spin-$s$ representation and $h_{ij}$ is the spin-$\half$
representation of $SU(2)$. This $SU(2)$ will form the $SU(2)_L \sim SO(3)_L$ in $SO(4)$.
A similar choice of states will be made for the tilde sector, 
with $SU(2)_R \sim SO(3)_R$ instead of $SU(2)_L\sim SO(3)_L$.

Since we do not consider excitations for the matter part, 
the intermediate states $\vert k\ra$ are of the form
$\vert \beta 0\ra$ and we can take
$U_{\beta\alpha} \equiv \la \beta 0\vert U \vert \alpha 0\ra$ to be a unitary matrix.
(Elements such as $\la \beta J\vert U \vert \alpha 0\ra$, $J \neq 0$, will correspond to transitions from
the matter vacuum to excited states of matter.) Further, we must count all states
in $\H_1 \otimes \H_2$, i.e., $P_+$ is the identity.
The action (\ref{36}) then simplifies as
\beq
S = - \int dt~ \left[\Tr (i A_0 )_L - \Tr (i A_0)_R\right]
\label{47a}
\eeq
Focusing on the left chirality part first, the states $\vert \alpha\ra$ which correspond to
 a spin-$s$ representation of
$SU(2)_L$, with $N = 2 s +1$, $s = n/2$, where $n$ is a positive integer,
can be taken as arising from the
quantization of $\mathbb{CP}^1$ viewed as a phase space, with the symplectic two-form
$n \, \omega_K$, where $\omega_K$ is the K\"ahler two-form on $\mathbb{CP}^1$.
In extracting the large $N$ limit, we can replace operators by their symbols which are the classical functions corresponding to them. 
If $\hat A$ is an operator acting on states of
the form $\vert \alpha , i\ra$, the symbol for $\hat A$ is defined as
\beq
({\hat A})_{ik} = A_{ik} = \la - s, i \vert \, h^{(s)\dagger} \,{\hat A}\, h^{(s)}\, \vert - s , k\ra
\label{48}
\eeq
where
$\vert -s \ra$ is the highest weight state of the spin-$s$ representation.
As a $2\times 2$ matrix, $h$ may be explicitly parametrized as
\beq
h = {1\over \sqrt{1 + \bz z} } \, \left( \begin{matrix} 1& z\\  -\bz & 1\\ \end{matrix}
\right) \, \left( \begin{matrix}  e^{i \theta/2} &0\\ 0& e^{-i \theta/2} \\ \end{matrix} \right)
\label{49}
\eeq
where $z, \bz$ are coordinates for one coordinate patch of $\mathbb{CP}^1 $. $h^{(s)}$ in
(\ref{48}) is the spin-$s$ representative of $h$ in (\ref{49}).
 Notice that the symbol is independent of $\theta$ and is a function on $\mathbb{CP}^1$.
It is also a $2\times 2$ matrix. 
We now define differential operators on $SU(2)$ by
\beq
R_a \, h = h\, t_a 
\label{50}
\eeq
where $t_a$, $ a = 1, 2, 3$, form a basis for the Lie algebra of $SU(2)$, 
$t_a = \sigma_a/2$ for the $2 \times 2$ representation. Explicitly in terms of the group parameters
$\vf^i$  (like $\bz, z, \theta$ in (\ref{49})),
\beq
R_a = i \, (E^{-1})^i_a {\del \over \del \vf^i} , \hskip .3in
h^{-1} \, dh = -i \, t_a\, E^a_i \, d\vf^i
\label{51}
\eeq
Because of the highest weight condition for the states in (\ref{48}),
$h \, \vert -s, k\ra$ obeys a holomorphicity condition,
\beq
(R_1 -i R_2 ) \, h \, \vert -s, k\ra =
R_- \, h\, \vert -s, k\ra = h\, t_- \vert -s, k\ra = 0
\label{52}
\eeq
The symbol or the classical function corresponding to an operator product ${\hat A}{\hat B}$ is
given by
\beqar
( {\hat A} {\hat B} )_{ik} &=&
\la -s, i\vert h^\dagger \, {\hat A} {\hat B} \, h \, \vert -s, k\ra\nonumber\\
&=& \sum_{a,j} \la -s, i\vert h^\dagger \, {\hat A}\, h \, \vert a, j\ra \la a, j \vert \, h^\dagger {\hat B} \, h \, \vert -s, k\ra\nonumber\\
&=& A_{ij} B_{jk} + \sum_{r=1}^{N-1} \la -s, i\vert h^\dagger \, {\hat A}\, h \, \vert -s+r, j\ra \la -s+r, j \vert \, h^\dagger {\hat B} \, h \, \vert -s, k\ra\nonumber\\
&=& A_{ij} B_{ik} + \sum_{r=1}^{N-1} \left[ {(N-1 -r)! \over r! (N-1)!}\right]\,(R_+ ^r \, A)_{ij} (R_-^r\, B)_{jk} 
\label{53}\\
&=& (A * B)_{ik}
\nonumber
\eeqar
The right hand side of this equation defines the $*$-product which starts off with the matrix product of the functions $A$, $B$ followed by additional terms involving derivatives.
As $N$ becomes large, these derivative terms are suppressed by powers of $N$.
Another useful result is that
the trace of an operator $\hat A$ can be expressed in terms of the integral of its symbol as
\beq
\Tr ({\hat A}) = \int d\mu \, \Tr \,A
\label{53a}
\eeq
The remaining trace on the right hand side is just over the matrix elements of the $2 \times 2$
matrix $A$.

Starting from the Hilbert space, the action
(\ref{36}) has a ``gauge invariance" corresponding to the transformation
\beq
U \rightarrow g \, U , \hskip .3in {\mathcal A} \rightarrow
g\, {\mathcal A}\, g^{-1} + \,{\dot g} \, g^{-1}
\label{41}
\eeq
For a gauge transformation with $g$ close to the identity, we can write
$g \approx 1 - \Phi$ and (\ref{41}) simplifies as
\beq
{\hat {\cal A}} \rightarrow {\hat {\cal A}} - \del_0 {\hat\Phi} 
- {\hat {\cal A}} \, {\hat\Phi}  + {\hat\Phi}\, {\hat {\cal A}},
\label{54}
\eeq
where $\del_0 = \del/\del t$ and we use the hat-notation to emphasize that all quantities are still operators.
We can represent (\ref{54}) in terms of the symbols as
\beq
{\cal A} \rightarrow {\cal A} - \del_0 \Phi - {\cal A} * \Phi
+ \Phi * {\cal A}
\label{55}
\eeq
This transformation still has the full content of the transformation at the level of operators.
The symbols ${\cal A}$ and $\Phi$ are functions of the coordinates of
$M = \mathbb{CP}^1 \times \mathbb{R}$ ($\mathbb{R}$ is for the time variable), and are also $2\times 2 $ matrices.
For an explicit realization of this transformation it is convenient to introduce a set of auxiliary quantities $A_\mu dx^\mu$ which is a one-form on $M$ such that
an ordinary continuum gauge transformation of $A_\mu$ with parameter
$\Lambda$ induces
the transformation (\ref{55}). 
In other words we seek functions
\beq
{\cal A}_0 =  {\cal A}_0 ( A_0, A_i) , \hskip .3in
\Phi = \Phi (\Lambda , A_0, A_i)
\label{56}
\eeq
such that
\beq
\left. \begin{matrix}
A_0 \rightarrow A_0 + \del_0 \Lambda + [A_0 , \Lambda ]\\
A_i \rightarrow A_i + \del_i \Lambda + \,[ A_i, \Lambda ]\\
\end{matrix} \right\} 
\Longrightarrow {\cal A}_0 + \del_0 \Phi + {\cal A}_0 * \Phi
- \Phi * {\cal A}_0
\label{57}
\eeq
The fact that this can be done is the essence of the Seiberg-Witten transformation
\cite{seiberg-witten}.
$A_0$ may be considered as the continuum version of $\A$ while $A_i$, the spatial
components, are additional auxiliary variables.
The choice of these $A_i$ is part of how the large $N$ limit is taken.
Our approach will be to optimize this choice by the equations of motion following from the large $N$ limit of the action.

The solution for (\ref{56}) is straightforward, although somewhat involved algebraically and
reads \cite{universalCS}
\beqar
{\cal A}_0 &=& A_0 + {P^{ab}\over 2 n} \left[
\del_a A_0 \, A_b - A_a \del_b A_0 + F_{a0} A_b - A_a F_{b0} \right] 
+  \cdots\label{58}\\
\Phi &=& \Lambda + {P^{ab}\over 2 n} ( \del_a \Lambda \, A_b - A_a \del_b \Lambda ) + \cdots\nonumber\\
P^{ab} &\equiv& {1\over 2}\left[ {g^{ab}\over 2 \pi} + i\, (\omega_K^{-1})^{ab}\right]
\nonumber
\eeqar
While we do not explicitly display the matrix labels, it should be kept in mind that these are all
$2\times 2$ matrices and matrix products are assumed. Taking the trace and using (\ref{53a}) 
we get 
\beq
\int dt~ \Tr {\cal A} = -{1\over 4\pi} \int \Tr \left[ (a + A) \, d (a+A) +{ 2\over 3} (a + A )^3 \right]
\label{59}
\eeq
where $a$ is the connection for $\omega_K$, i.e., $\omega_K = d a$. The result is the integral of 
a Chern-Simons term, expanded around $a$ as a background. It is useful to think of $a +A$ as the connection of interest, in this case $SU(2)_L$-valued. The dependence on the choice of
$\mathbb{CP}^1$ with its $\omega_K$ as a background to expand around is irrelevant once we get to
(\ref{59}). The background is thus only an auxiliary step in arriving at this result.
There are higher terms, with more derivatives and so on, which do retain the dependence
on the metrical details of the background. These are negligible for slowly varying
$A_\mu$; some of the terms will also cancel out when we include the $SU(2)_R$ sector.

The final result of simplifying (\ref{36}) would thus be
\beq
S = - {1\over 4\pi} \int \left[ \Tr \left( A \, dA + {2\over 3} A^3 \right)_L - 
\Tr \left( A \, dA + {2\over 3} A^3 \right)_R \right]
\label{60}
\eeq
The $A$'s are connections for $SU(2)_L$ and $SU(2)_R$ and this result 
reproduces the Euclidean gravitational action (\ref{43}), apart from an overall
factor of $(l/8\,G)$. This overall degeneracy factor is important and, in this approach, has to come from several copies of the fields we have used.

We have used the action (\ref{36}) as it makes more transparent the role of the spatial
components of $A$ as part of how the large $N$ limit is taken.
We can also obtain the result (\ref{60}) from the action 
(\ref{new18}) by integrating out the spinor fields.
In the present case, we have $\M = {\mathbb{CP}}^1$ is two-dimensional.
We take the fermions to be in the spin-$\half$ representation of
$SU(2)$, i.e., the indices $I, J$ takes values $1, 2$. 
The fermions $\Psi$ are coupled to
the $SU(2)_L$ gauge fields while $\Phi$ couple to $SU(2)_R$ fields.
We have a field theory of Dirac fermions in 2+1 dimensions and it is well known that
the effective cation, upon integrating out the fermions, is the Chern-Simons action.
Since the fields $\Phi$ are defined on $\M$ with the orientation reversed, the Chern-Simons action generated by these fields will be the negative of the one generated by $\Psi$.
The result, once again, is the action (\ref{60}).
In fact, being a field theory, the action (\ref{new18}) is much easier to use in practice as there are familiar diagrammatic techniques for evaluating the effective action.

We have considered the Euclidean signature so far. The spinor field version of the action, namely,
(\ref{new18}), can be easily continued to Minkowski signature, with the gauge fields
being valued in the Lie algebra of $SO(2,1)_L $ and $SO(2,1)_R$.

\section{Discussion}

A number of comments are in order at this point.

We have carried out the calculations in section 3 for Euclidean signature. While this is not relevant for the differential form, the trace in the Chern-Simons action over the gauge fields which are written as matrices corresponding to the spin-$\half$ representation of $SU(2)_L$ and $SU(2)_R$
is sensitive to the signature.
A continuation to Minkowski signature would involve $SO(2,1)$ representations, which in the interest of unitarity, should be chosen as infinite dimensional.
Defining the trace is then rather tricky. However, the formulation of TFD in terms of the
fermionic field theory, as in (\ref{new17}), (\ref{new18}) avoids this problem. 
Taking the fields in (\ref{new18}) to be $SO(2,1)$ spinors, we have a standard fermionic field theory 
and we expect that this will be consistent with unitarity.

There are many papers analyzing 2+1 dimensional gravity, starting with the Chern-Simons formulation and considering the evaluation of the partition function \cite{{witten}, {maloney}, {2+1others}}.
All these lead to strong hints of the underlying string origin of the action. This may very well be the case, but since we are obtaining the Chern-Simons action only in the large $N$ limit, 
it is not clear how to compare our work with these developments.
However, we may note the following. We have used a rank $n$ representation of $SU(2)$ with
$n \rightarrow \infty$ to approximate the two-dimensional spatial manifold.
In Minkowski signature, we should presumably use a representation of $SO(2,1)$.
But we may also consider using a coadjoint orbit of the Virasoro group 
(which contains a suitable $SO(2,1) \sim SL(2, {\mathbb R})$) to construct
a noncommutative version of the two-manifold \cite{witten2}.
 In that case,
the analogue of the large $N$ limit would be the limit of large central charge. Indeed this is the limit considered in
\cite{maloney} to compare 2+1 dimensional gravity and string theory. 
Further elaboration of this connection has to be left to future work at this stage.

Finally, we may note that the spectral action of Connes \cite{connes} is not the same as  what we have, but there is some similarity. 
A key result for the spectral action
 is that the Wodzicki residue of the inverse square of the Dirac operator
gives the Einstein action, see in particular Kastler's article in \cite{connes}.
In our case, we are considering odd dimensional spacetimes. However, the action is basically
related to the Dirac action given in (\ref{new18}). The main points of difference are that
we have two sets of spinor fields which are naturally obtained in
thermofield dynamics, and that they carry opposite chiralities in terms of coupling to gravitational
degrees of freedom.
Further, we are evaluating the effective action in a limit
corresponding to $c \rightarrow \infty$ or 
we are taking the limit where only the lowest modes of the Dirac Hamiltonian
of (\ref{new18}) is included.
Whether this is related in some fashion to the Wodzicki residue is not clear.

Another point of clarification is the following.
There are two ways to think of gravity on noncommutative spaces. We may consider the 
continuum description as an approximation, with a noncommutative
operator version at short distances or at finite $N$. 
In this case, one has the full set of dynamical fields
for gravity at all levels \cite{connes}-\cite{steinacker}.
An alternative is to consider gravity
as being trivial at the fundamental level only emerging in the continuum limit. This would be
more in the spirit of gravity as an entropic phenomenon \cite{jacobson}-\cite{verlinde}.
In our case, the
 spatial components of the gauge field were introduced in (\ref{57}) to provide a simple
realization of the gauge transformation property (\ref{55}), with their values set to what is given by extremizing the continuum action. 
Thus there is no dynamics for these components if we stay at the level of the starting action
(\ref{47a}) or (\ref{36}), making this approach closer to 
\cite{jacobson}-\cite{verlinde}.

\bigskip
I thank A.P. Balachandran and Alexios Polychronakos for useful comments and criticism.
This research was supported by the U.S.\ National Science
Foundation grant PHY-1213380
and by a PSC-CUNY award. 

\section*{Appendix: A generalization of (\ref{34})}
\def\theequation{A\arabic{equation}}
\setcounter{equation}{0}

The action for the functional integral for thermofield dynamics was given
in (\ref{34}) for a single quantum system. The fermion field version of that was given 
in (\ref{new16}) and generalized to a multipartite system with identical
components in (\ref{new17}, \ref{new18}).
Here we want to consider the direct generalization of (\ref{34}) and its relation to the
Hall effect.

if we consider several systems which are distinct, the generalization is simple.
We must have a unitary matrix $U$ for each system and
the action is simply a sum of actions of the form (\ref{34}).
The more interesting case is when we have identical subsystems; each 
subsystem may be 
a boson or a fermion. The latter is what is relevant for noncommutative geometry.
In that case, we have a Hilbert space $\H$, say, of dimension $N$. States in this $\H$ replace the notion of points on a manifold. All of $\H$ must be included to account for the total volume of the
manifold. For the purpose of time-evolution, it is then useful to think of
the space as a state in the $N$-fold tensor product $\H \otimes \H \cdots \otimes \H$.
To avoid double counting, the choice of state in each component $\H$ must be distinct.
The simplest way to implement this is to take the totally antisymmetrized state in 
$\H \otimes \H \cdots \otimes \H$, i.e., an $N$-fermion state.

An analogy with a quantum Hall system is very useful for this situation.
Consider
a quantum Hall system
in the lowest Landau level with one-particle states
$\vert \alpha \ra$, $\alpha = 1, 2, \cdots, N$.
To be general, we assume that the fermion can have multiple internal states, such as spin degrees of freedom. (In the context of noncommutative geometry, this could correspond to degrees of freedom
of matter.)
We use labels $I, \, J$, etc. for the latter, so that states may be represented as
$\vert k \ra = \vert \alpha \, I\ra$ which span a Hilbert space $\H$.
Let $M$ be the dimension of this space.
 We consider the dynamics of a droplet of $n$ fermions starting out in
 one spin state, say, $I = 0$.
The many-body Hilbert space
will consist of suitably antisymmetrized states in the $n$-fold tensor
product $\H \otimes \H \otimes \cdots\otimes \H$. 
Let ${\cal N}$ denote the dimension of this Fock space,
${\cal N} = M!/((M-n)! n! )$.
Let us say that we are considering the first $n$ states,
i.e., $\vert \alpha\, 0\ra$, $\alpha = 1, 2, \cdots, n$, to be filled.
This means that if we write a fermionic field operator
$\psi (x) = \sum_{k}  a_{k } \, f_{k } (x)$,
where $f_{k }(x)$ are the single particle wave functions,
the droplet is given by
the state $a_{1 0}^\dagger \, a_{2 0}^\dagger \cdots a_{n 0}^\dagger \vert 0\ra$
or by the density matrix 
\beq
\rho = a_{10}^\dagger \, a_{20}^\dagger \cdots a_{n 0}^\dagger \vert 0\ra \,\la 0\vert 
a_{n 0}  \, \cdots a_{2 0}  \, a_{1 0}
\label{36b}
\eeq
The possible transformations of this configuration, including time-evolution,
 are given by an element of $U ({\cal N})$.
 We have a single pure state chosen by $\rho$ in (\ref{36b}), so
a path integral for the time-evolution of this state would have the action
 (\ref{7}), with integration over all $ {\cal U}(t) \in U({\cal N})$
 modulo transformations which leave $\rho$ invariant. 
 The relevant action is thus
 \beqar
 S &=& \int dt\, \Bigl[  \sum_{\{k \} }  i\,
 ({\cal U}^\dagger )_{10\,20\,\cdots n 0, k_1 k_2 \cdots k_n}
  (\dot {\cal U})_{k_1 k_2 \cdots k_n, 10\, 20\,\cdots n 0} \nonumber\\
  &&\hskip .85in - \sum_{\{ k l\}}
   ({\cal U}^\dagger )_{10\,20\,\cdots n 0, k_1 k_2 \cdots k_n} \, H_{k_1 k_2 \cdots k_n, \,
   l_1 l_2\cdots l_n} 
  ({\cal U})_{l_1 l_2 \cdots l_n, 10\,20\,\cdots n 0}\Bigr]
  \label{36c}
  \eeqar
This is the general situation for the case with all sorts of many-body interactions.
 However, if the Hamiltonian and other observables
 of interest only involve singe-particle operators, i.e., they are of the
 form $C = a^\dagger_k \, C_{k l} \, a_l$, then the action simplifies. The possible
 transformations will correspond to unitary transformations
 of the form $U_{ k l} \in U(M)$ acting on the single particle states,
 $U$ being of the form $e^{i C}$.
 The states in the $n$-body Hilbert space
 may be viewed as corresponding to an irreducible representation
 of the group $U(M)$ obtained as the totally antisymmetrized product of 
 $n$ copies of the fundamental
 representation of $U(M)$. Thus we can simplify (\ref{7}) by restricting to
 $U(M) \in U ({\cal N})$. In this case the elements of
 ${\cal U}$ are given by the Slater determinant,
 \beq
   ({\cal U})_{k_1 k_2 \cdots k_n, 10\,2 0\, \cdots n 0}
   = {1\over \sqrt{n!}}~\left| \begin{matrix} 
   U_{k_1\,1 0} &   U_{k_1\,2 0} &\cdots&    U_{k_1\,n 0} \\
      U_{k_2\,1 0} &    U_{k_2\,2 0} &\cdots&    U_{k_2\,n 0} \\
      \cdots\\
      \end{matrix}
      \right|
 \label{36d}
\eeq
It is easily seen that the action (\ref{36c}) simplifies to
\beqar
S &=& \int dt~ \Tr \left[ P_+\, \left( i\, U^\dagger \,  {\dot U} - U^\dagger \, H \, U
\right) \right]\nonumber\\
&=& \int dt~ \sum_{\alpha = 1}^n \left[i\,  \la \alpha \, 0\vert U^\dagger \vert k \ra\, \la  k \vert {\dot U} \vert 
\alpha\,0\ra
 - \la \alpha \, 0 \vert U^\dagger \vert  k \ra \, H_{k l} \, \la l \vert U \vert \alpha \, 0\ra
 \right]\label{38}\\
P_+ &=& \sum_{\alpha = 1}^n \, \vert \alpha 0\ra \la \alpha 0\vert \label{38c}
\eeqar
Notice that $P$ is not a density matrix; 
it is a representation, at the level of the one-particle Hilbert space, of the density matrix
for the pure state (\ref{36b}) of the droplet of $n$ fermions.
(It was this version of the action which was used in
\cite{{sakita}, {Nair1}}.)
In this expression, we do not yet have the part with the negative eigenvalues of $P$ 
since the tilde system is not included. It is easy to see that 
the tilde system will have a similar path integral, with the eigenvalues of
$P$ being $-1$. The complete action for the thermofield dynamics of this system is
then
\beqar
S &=& \int dt~ \sum_{\alpha = 1}^n \left[i\,  \la \alpha \, 0\vert U^\dagger \vert k \ra\, \la  k \vert {\dot U} \vert 
\alpha\,0\ra
 - \la \alpha \, 0 \vert U^\dagger \vert  k \ra \, H_{k l} \, \la l \vert U \vert \alpha \, 0\ra
 \right]\nonumber\\
&& - \int dt~ \sum_{\alpha=1}^n
 \left[i\,  \la \alpha \, 0\vert {\tilde U}^\dagger \vert k \ra\, \la  k \vert {\dot {\tilde U}} \vert 
\alpha\,0\ra
 - \la \alpha \, 0 \vert {\tilde U}^\dagger \vert  k \ra \, H_{k l} \, \la l \vert {\tilde U} \vert \alpha \, 0\ra
 \right]
 \label{36}
\eeqar
When applying this to noncommutative geometry, we should take
$n =N$.

Here is a curiosity in this formulation:
In the path integral, we also have the functional integration over
$\mathbb{CP}^{{\cal N}-1}$; this corresponds to the integration
over $ ({\cal U})_{k_1 k_2 \cdots k_n, 10\, 20\,\cdots n 0}$ at each instant of time.
In restricting the action to $U(M)$ transformations with
$P$ as in (\ref{38}), a part of this integration becomes trivial.
The left over integration is over the Grassmannian space
$U(M)/ U(n)\times U(M-n)$. This gives a factor
of $V$ at each instant of time, where $V$ is defined by
\beq
\int d\mu (\mathbb{CP}^{{\cal N}-1} ) = V~ \int d\mu (U(M)/ U(n)\times U(M-n) )
\label{38a}
\eeq
Exponentiating these factors we get an additional term in $S$ which is
\beq
\Delta S = \int {dt \over \epsilon} \, \log V
\label{38b}
\eeq
Here we are considering dividing the interval of time-integration into segments
of length $\epsilon$, with $\epsilon \rightarrow 0$ eventually.
This extra factor may be thought of as an additional entropy factor arising from the fact that we are 
restricting the observables to a smaller set, namely, to those of the one-particle type.
(It is also similar to the terms with $\delta (0)$-factor that one encounters in field theories
when making a change of field variables upon exponentiating the
Jacobian of the transformation.)

After this discussion, it should be clear that (\ref{36}), apart from the extra term
(\ref{38b}), is indeed the required generalization once
the tilde system is included.

This generalization to several systems can also be phrased as 
a contour integral, with the action
\beq
S_C = i\,\oint_C dt~\sum_{\alpha =1}^n \left(\,U^\dagger  {\dot U} - U^\dagger \A \,U \right)_{\alpha 0\, \alpha 0} 
\label{40a}
\eeq
%%%%%%%%%%%%%%%%%%%%%%%%%%%%%%%%%%%%%%%%%%%%%%%%
%%%%%%%%%%%%%%%%%%%%%%%%%%%%%%%%%%%%%%%%%%%%%%%%

%%%%%%%%%%%%%%%%%%%%%%%%%%%%%%%%%%%%%%%%%%%%%%%%
%%%%%%%%%%%%%%%%%%%%%%%%%%%%%%%%%%%%%%%%%%%%%%%%
%%%%%%%%%%%%%%%%%%%%%%%%%%%%%%%%%%%%%%%%%%%%%%%%
%%%%%%%%%%%%%%%%%%%%%%%%%%%%%%%%%%%%%%%%%%%%%%%%
\end{document}